\documentclass[aps,showpacs,twocolumn]{revtex4}
\usepackage{epsfig}
\usepackage{slashed}
\usepackage{color}

\begin{document}

\title{QCD phase diagram at finite isospin and baryon chemical potentials with the self-consistent mean field approximation}

\author{Zu-Qing Wu$^a$}\email{wujieyi1001@foxmail.com}
\author{Jia-Lun Ping$^b$}\email{jlping@njnu.edu.cn}
\author{Hong-Shi Zong$^{a,c,d}$}\email{zonghs@nju.edu.cn}

\affiliation{$^a$Department of Physics, Nanjing University, Nanjing 210093, P. R. China}

\affiliation{$^b$Department of Physics, Nanjing Normal University, Nanjing 210023, P. R. China}

\affiliation{$^c$Nanjing Proton Research and Design Center, Nanjing 210093, P. R. China}

\affiliation{$^d$Department of Physics, Anhui Normal University, Wuhu 241000, P. R. China}

\begin{abstract}
The self-consistent mean field approximation of two-flavor NJL model with introducing a free parameter $\alpha$
to reflect the competition between "direct" channel and the "exchange" channel, is employed to study QCD phase structure
at finite isospin chemical potential $\mu_I$, finite baryon chemical potential $\mu_B$ and finite temperature $T$,
especially the location of the QCD critical point. It is found that, for fixed isospin chemical potentials the lower temperature of phase transition is obtained with $\alpha$ increasing in the $T-\mu_I$ plane, and the largest difference of the phase transition temperature with different $\alpha$'s appears at $\mu_I \sim 1.5m_{\pi}$. At $\mu_I=0$ the temperature of the QCD critical end point (CEP) decreases with $\alpha$ increasing, while the critical baryon chemical potential increases. At high isospin chemical potential ($\mu_I=500$ MeV), the temperature of the QCD tricritical point (TCP) increases with $\alpha$ increasing, and in the regions of low temperature the system will transit from pion superfluidity phase to the normal phase as $\mu_B$ increases. At low temperatures, the critical temperature of QCD phase transition with different $\alpha$'s rapidly increases with $\mu_I$ at the beginning, and then increases smoothly around $\mu_I>300$ MeV. In high baryon density region, the increase of the isospin chemical potential will raise the critical baryon chemical potential of phase transition.
\end{abstract}


\maketitle

\section {Introduction}
As is known that at low temperature and low baryon chemical potential the strongly interacting system is in hadronic phase, and
as temperature and/or baryon chemical potential increase the system will transit to quark-gluon plasma (QGP) phase in which the quarks
and gluons are deconfined and chiral symmetry is partially restored. This new state of matter can be created from a hot and dense fireball,
which is able to reach the transition temperature, in high energy nucleus-nucleus collisions \cite{P}. The early Universe should have been
in this phase for the first few microseconds after the Big Bang, and the QGP is also expected in the interior of quark stars or hybrid
stars \cite{Sineeba}. For recent twenty years, the study on quantum chromodynamics (QCD) phase diagram is extended to finite isospin chemical
potential \cite{Bastian,Xuanmin,Juliane}. The physical motivation to study QCD at finite isospin chemical potential is related to
the investigation of neutron stars, isospin asymmetric nuclear matter, and heavy ion collisions using neutron-rich heavy-ion beams.
The isospin chemical potential has an effect on hadronic matter, which can rotate the quark-antiquark condensation, and the phenomenon is
called as pion condensate, because it indicates the direction of the $U_I(1)$ symmetry breaking related to the conservation of the pion number.
When the isospin chemical potential exceeds the pion mass ($\mu_I>m_\pi$), a superfluid of charged pions will appear in the zero momentum state,
i.e. the pion superfluidity. This Bose-Einstein condensation (BEC) of pions is an electromagnetic superconductor \cite{Stefano,D.T.Son}.
Distinguished from the normal phase ($\mu_I \le m_\pi$) where pion condensate equals to zero, the realization of pion condensate can change
the low energy properties of matter, such as the lifetimes and the mass spectrum of mesons \cite{Andrea,Migdal,Kogut}, and it also relates
to lots of phenomena \cite{B,Viktor}. Therefore, it is important for us to study the phase transition of strongly interacting system in baryon
and isospin medium.

It is generally thought that QCD is viewed as the fundamental theory of strong interaction. Theoretically, there exist rich phase structures of
QCD at finite temperature and finite density. In the region of high temperatures and/or high densities, the perturbative QCD can work well
to the properties of the phase structures. Finite temperature region with vanishing baryon chemical potential, lattice simulations have provided valuable insights into the QCD phase diagram. Although lattice simulation are hindered by the "sign" problem when dealing with finite chemical potential, it can in principle deal with the problem of finite isospin  chemical potential \cite{Karsch,Son}. Furthermore, there are a lot of low-energy effective models, such as the chiral perturbation theory \cite{Loewe,Kogut}, random matrix method \cite{Klein,Arai},
quark-meson model \cite{Anthony,Bernd-Jochen} and Nambu--Jona-Lasinio (NJL) model \cite{Barducci,He,Tao,Lian}, to be used as tools to investigate
the phase structures in isospin and baryon matter, in which the NJL model describes the chiral dynamics of QCD well \cite{S.P.Klevansky}.

In this paper, we study QCD phase diagram at finite isospin and baryon chemical potentials in the framework of NJL model with self-consistent
mean field approximation \cite{FeiWang}. The standard Lagrangian of NJL model contains scalar $\left(\bar{\psi}\psi\right)^2$ and
pseudoscalar-isovector $\left(\bar{\psi}i\gamma_5\mbox{\boldmath{$\tau$}}\psi\right)^2$ channels \cite{S.P.Klevansky}, which is insufficient
for describing physics in the case of finite density. Because of this, people often use the mathematical identity transformation -- Fierz
transformation to generate other interaction channels. These interaction terms play an important role in the the case of finite density.
For instance, when we discuss the finite chemical potential, the vector-isoscalar channel is very important \cite{Walecka}.
Similarly when the axial chemical potential is involved, the isovector-isoscalar channel becomes very important \cite{S,BW,YL,ZF},
and in the study of chirally imbalanced systems, the contribution of the axial-vector channel cannot be neglected \cite{Yang}.
In this connection, if we study the system under the conditions of finite isospin and baryon densities, the contributions of the
vector-isoscalar channels and the pseudoscalar-isovector channels should be considered \cite{Z}. In previous analyses of NJL model,
people usually ignore the contributions of various channels from the Fierz-transformed term or add the relevant terms by hand \cite{Kunihiro}.
As shown below, the above mean field approximation approach is not self-consistent. In this paper, we will employ the self-consistent
mean field approximation \cite{FeiWang} of NJL model to study the phase structure of strongly interacting system at nonzero isospin and baryon
chemical potentials. This model introduces a free parameter $\alpha$ to reflect the proportion of the different channel contributions from
the Fierz-transformed term.

This paper is organized as follows. In Sec. \ref{s2} the two-flavor NJL model and the self-consistent mean field approximation
in the finite isospin and baryon chemical potentials are introduced and we can get the self-consistent gap equations. In Sec. \ref{s3}, we discuss
QCD phase structure, and analyze our numerical results with different $\alpha$'s. In the last section, we summarize our findings.

\section {The self-consistent mean field approximation of NJL model}
\label{s2}
In the present study, only two flavors light quarks are considered, i.e., $N_f=2$. The standard two-flavor NJL model
Lagrangian density is defined as \cite{S.P.Klevansky}
\begin{equation}
{\cal L}_{NJL} =
\bar{\psi}\left(i\slashed{\partial}-m_0\right)\psi
+G\left[\left(\bar{\psi}\psi\right)^2+\left(\bar{\psi}i\gamma_5\mbox{\boldmath{$\tau$}}\psi\right)^2
\right]
\end{equation}
with scalar and pseudoscalar interactions corresponding to $\sigma$ and ${\bf \pi}$ excitations separately, where
$m_0 = \text{diag}(m_{0u},m_{0d})$ is the matrix of the current quark mass and the coupling constant is $G$.

Performing the Fierz transformation \cite{S.P.Klevansky} on the four-Fermion interaction terms, one has
\begin{eqnarray}
&{\cal L}_{IF}& =
\frac{G}{8N_c} \left[2\left(\bar{\psi}\psi\right)^2+2\left(\bar{\psi}i\gamma_5\mbox{\boldmath{$\tau$}}\psi\right)^2
-2\left(\bar{\psi}\mbox{\boldmath{$\tau$}}\psi\right)^2 \right. \nonumber\\
&& -2\left(\bar{\psi}i\gamma_5\psi\right)^2-4\left(\bar{\psi}\gamma^{\mu}\psi\right)^2-4\left(\bar{\psi}i\gamma^{\mu}\gamma_5\psi\right)^2
 \nonumber\\
&&\left.+\left(\bar{\psi}\sigma^{\mu\nu}\psi\right)^2 -\left(\bar{\psi}\sigma^{\mu\nu}\mbox{\boldmath{$\tau$}}\psi\right)^2
\right],
\end{eqnarray}
where the contributions of color octet have been neglected and $N_c=3$ is the number of colors. Then the Lagrangian becomes
\begin{equation}
{\cal L}_F =
\bar{\psi}\left(i\slashed{\partial}-m_0\right)\psi+{\cal L}_{IF}.
\end{equation}

The original Lagrangian ${\cal L}_{NJL}$ and the transformed Lagrangian ${\cal L}_{F}$ are equivalent, since the Fierz transformation
is a mathematical identity transformation. In view of the ${\cal L}_{NJL}$ and ${\cal L}_{F}$ are mathematically equivalent, the most
general effective Lagrangian can be introduced \cite{FeiWang}: ${\cal L}_R = (1-\alpha) {\cal L}_{NJL} + \alpha {\cal L}_F$, where
the parameter $\alpha$ is a weighting factor (a real number from 0 to 1) used to reflect the competition between "direct" channel
(${\cal L}_{NJL}$) and the "exchange" channel (${\cal L}_{F}$).
In this way, we can obtain more general interaction channels through Fierz transformation. It helps us to understand and deal with
the problem of strongly interacting system with finite density. As mentioned in the last section, the various channel contributions
are not negligible if we study the system in the case of finite density.

Nevertheless, once the mean field approximation is applied, the contributions of ${\cal L}_{NJL}$ and ${\cal L}_{F}$ are no longer
identical, because the Fierz transformation and the mean field approximation are not commutative. Especially when the system is at
finite density, the results yielded by the two Lagrangians are very different \cite{S.P.Klevansky}. This means that it is important
for us to know the contributions of each interacting channel once the mean field approximation is used. In fact, just as pointed out
by Refs. \cite{FeiWang,TongZhao,Q,Yang,Z,C,Zi}, there is no physical requirement to determine the value of $\alpha$ currently. In principle,
$\alpha$ needs to be constrained by experiments rather than the self-consistent mean field approximation itself. The Lagrangian under
the self-consistent mean field approximation is adopted as
$\langle{\cal L}_R\rangle_m = \left(1-\alpha\right) \langle{\cal L}_{NJL}\rangle_m + \alpha \langle{\cal L}_F\rangle_m$ \cite{FeiWang},
where $\langle ¡­\rangle_m$ denotes the mean field approximation. What needs to be emphasized here is that the mean field approximation
that dose not consider the contribution of the "exchange" channel (${\cal L}_{F}$) is theoretically not self-consistent \cite{T.Kunihiro,T.Hatsuda}.

In order to investigate the system in isospin and baryon matter, we can introduce the isospin chemical potential $\mu_I$ and the baryon
chemical potential $\mu_B$, which connect to the isospin number density $n_I=(n_u-n_d)/2$ and the baryon number density $n_B=(n_u+n_d)/3$
respectively. In the imaginary time formulism of finite temperature field theory \cite{J.I.Kapusta}, the partition function for a system
at finite baryon and isospin densities can be represented as
\begin{equation}
\label{s2 40}
Z\left(T,\mu_I,\mu_B,V\right) = \int \left[d\bar{\psi}\right]\left[d\psi\right]
  e^{\int_{0}^{\beta} d\mbox{\boldmath{$\tau$}} \int d^3\vec{x} \left({\cal L}
 + \bar{\psi}\mu\gamma_0\psi\right)}\ ,
\end{equation}
where $V$ is the volume of the system, $\beta = 1/T$ is the inverse temperature; $\mu_I$ and $\mu_B$ are the isospin and baryon
chemical potentials, where $\mu = \text{diag}(\mu_u,\mu_d)$ is the matrix of quark chemical potential in flavor space with the $u$
and $d$ quark chemical potentials,
\begin{eqnarray}
\label{s2 2}
\mu_u &=& \frac{\mu_B}{3}+\frac{\mu_I}{2}\ ,\nonumber\\
\mu_d &=& \frac{\mu_B}{3}-\frac{\mu_I}{2}\ ,
\end{eqnarray}
the factors $\frac{1}{2}$ and $\frac{1}{3}$ reflect the fact that quark's isospin quantum number is $\frac{1}{2}$ and $3$ quarks make up a baryon.

Then the equivalent Lagrangian can be rewritten as
\begin{equation}
{\cal L}_r= (1-\alpha) {\cal L}_{NJL} + \alpha {\cal L}_F +  \bar{\psi}\mu\gamma_0\psi.
\end{equation}

In our work, we only care about the contributions from scalar, vector and pseudoscalar-isovector channels. Other terms have no effects in
our calculation at the level of mean field approximation. Applying the mean field approximation to this Lagrangian, and dropping the irrelevant
terms, we can get the effective Lagrangian
\begin{eqnarray}
\label{s2 1}
{\cal L}_{eff}
&=&\bar{\psi}\left(i\slashed{\partial}-M+\mu^{\prime} \gamma_0
+2G\pi i\gamma_5\tau_1\right)\psi \nonumber\\
&-& G\left(\sigma^2 + \pi^2\right) +\beta n^2\ ,
\end{eqnarray}
where $M$ is called constituent quark mass:
\begin{equation}
\label{s2 9}
M=m_0-2G\sigma,
\end{equation}
and
\begin{equation}
\label{s2 3}
\mu^{\prime} = \mu - 2 \beta n.
\end{equation}

Employing the Eqs. (\ref{s2 2}) and (\ref{s2 3}), one can obtain the following relations,
\begin{equation}
\label{s2 10}
\mu^{\prime}_I = \mu_I - 8 \beta n_I,
\end{equation}
and
\begin{equation}
\label{s2 11}
\mu^{\prime}_B = \mu_B - 18 \beta n_B,
\end{equation}
where for convenience we redefine the parameter in the formalism,
$$\beta=\frac{-2 G \alpha}{11 \alpha-12}.$$

The quark condensation $\sigma=\langle\bar{\psi}\psi\rangle$, the pion condensation
$\pi=\langle\bar{u} i\gamma_5 d\rangle+\langle\bar{d} i\gamma_5 u\rangle$, the quark number density
$n=\langle\bar{u}\gamma_0u\rangle+\langle\bar{d}\gamma_0d\rangle$, the isospin number density
$n_I={1 \over 2}(\langle\bar{u}\gamma_0u\rangle-\langle\bar{d}\gamma_0d\rangle)$, and the baryon number density $n_B={1 \over 3}(\langle\bar{u}\gamma_0u\rangle+\langle\bar{d}\gamma_0d\rangle)$ can be determined in a thermodynamically self-consistent way.
We can insert the effective Lagrangian (\ref{s2 1}) into the partition function (\ref{s2 40}) to get the mean-field thermodynamic potential
\begin{eqnarray}
\Omega &=& -\frac{T}{V} \ln{Z} \nonumber\\
&=& G\left(\sigma^2 + \pi^2\right) - \beta n^2 + \Omega_M ,
\end{eqnarray}
where $\Omega_M$ is expressed as
\begin{eqnarray}
&& \Omega_M = -N_c \int_{0}^{\Lambda} {d^3\vec{p}\over (2\pi)^3} \left[E_-^- - E_+^- + E_-^+ - E_+^+ + 2T\right.\nonumber\\
&& \left(\ln\left(1+\exp{(-E_-^-/T)}\right) + \ln\left(1+\exp{(E_+^-/T)}\right) \right. +  \nonumber\\
&& \left.\left. \ln\left(1+\exp{(-E_-^+/T)}\right) +\ln\left(1+\exp{(E_+^+/T)}\right)\right)\right].
\end{eqnarray}
Here the effective quark energies $E^\pm_\mp$ are given by
\begin{eqnarray}
E^\pm_\mp &=& E^\pm_p \mp {\mu^{\prime}_B \over 3},\\
E^\pm_p &=& \sqrt{\left(E_p \pm \mu^{\prime}_I/2\right)^2 + 4G^2\pi^2} ,\\
E_p &=& \sqrt{|\mathbf{p}|^2 + M^2}\ .
\end{eqnarray}

Given the extremum condition of the thermodynamic potential $\frac{\partial{\Omega}}{\partial{\sigma}}=0,\frac{\partial{\Omega}}{\partial{\pi}}=0,\frac{\partial{\Omega}}{\partial{n}}=0
,\frac{\partial{\Omega}}{\partial{n_I}}=0,\frac{\partial{\Omega}}{\partial{n_B}}=0$, we can get the quark condensate,
\begin{eqnarray}
\label{s2 4}
\sigma &=& \int_{0}^{\Lambda} {d^3\vec{p}\over (2\pi)^3} \frac{2N_cM}{E_p} \left[\frac{E_p-\mu^{\prime}_I/2}{E_p^-}(f(E_-^-)-f(-E_+^-)) \right.\ \nonumber\\
&+& \left. \frac{E_p+\mu^{\prime}_I/2}{E_p^+}(f(E_-^+)-f(-E_+^+))\right]\ ,
\end{eqnarray}
the pion condensate,
\begin{eqnarray}
\label{s2 5}
\pi    &=& -4N_cG\pi \int_{0}^{\Lambda} {d^3\vec{p}\over (2\pi)^3} \left[\frac{1}{E_p^-} \left(f(E_-^-)-f(-E_+^-)\right) \right.\ \nonumber\\
&+& \left. \frac{1}{E_p^+} \left(f(E_-^+)-f(-E_+^+)\right)\right]\ ,
\end{eqnarray}
the quark number density,
\begin{eqnarray}
\label{s2 6}
n      &=& 2 N_c\int_{0}^{\Lambda} {d^3\vec{p}\over (2\pi)^3} \left[f(E_-^-)+f(-E_+^-)+f(E_-^+) \right.\ \nonumber\\
&+& \left. f(-E_+^+)-2\right]\ ,
\end{eqnarray}
the isospin number density,
\begin{eqnarray}
\label{s2 7}
n_I    &=& N_c\int_{0}^{\Lambda} {d^3\vec{p}\over (2\pi)^3} \left[ \frac{E_p-\mu^{\prime}_I/2}{E_p^-}(f(E_-^-)-f(-E_+^-)) \right.\ \nonumber\\
&-& \left. \frac{E_p+\mu^{\prime}_I/2}{E_p^+}(f(E_-^+)-f(-E_+^+))\right]\ ,
\end{eqnarray}
and the baryon number density,
\begin{eqnarray}
\label{s2 8}
n_B      &=& {2 \over 3} N_c\int_{0}^{\Lambda} {d^3\vec{p}\over (2\pi)^3} \left[f(E_-^-)+f(-E_+^-)+f(E_-^+) \right.\ \nonumber\\
&+& \left. f(-E_+^+)-2\right]\
\end{eqnarray}
with the Fermi-Dirac distribution function
\begin{equation}
f(x) = {1\over e^{x/T}+1}.
\end{equation}
Finally, inserting Eqs. (\ref{s2 4}-\ref{s2 8}) into the Eqs. (\ref{s2 9}-\ref{s2 11}), we will obtain a set of combined and compled integral
equations in the case of finite isospin and baryon chemical potentials. By numerically solving this set of equations we can get the relevant phase diagram.

The parameters used in the present work are the current quark mass $m_{0u}=m_{0d}=m_0=4.76$ $\text{MeV}$,
the cutoff $\Lambda = 659$ $\text{MeV}$, and the coupling constant $G = 4.78\times 10^{-6} \text{MeV}^{-2}$, which are obtained by fitting
the pion mass $m_{\pi}=131.7$ $\text{MeV}$ as used by lattice QCD \cite{Brandt} at $T = \mu_I = \mu_B = 0$, and other parameters are
the decay constant $f_{\pi}=92.4$ $\text{MeV}$ and the quark condensation per flavor $\langle\bar{\psi}\psi\rangle=-(250~\text{MeV})^3$.

\section {Numerical results and discussion} \label{s3}
It is mentioned above that the Refs. \cite{FeiWang,TongZhao,Q,Yang,Z,C,Zi} indicate the parameter $\alpha$ should be constrained by experiments.
A possible choice, for instance in Refs. \cite{TongZhao,Q}, is that $\alpha$ can be determined by astronomical observation data of the recent neutron
star merging. However, with the lack of reliable experiment data of the strongly interacting matter at finite densities currently, so in our work
we consider $\alpha$ as a free parameter. In this paper, we will show our results with different $\alpha$'s. $\alpha$=0 represents the standard NJL
model \cite{Lian,S.P.Klevansky}, $\alpha$=0.5, which is in good agreement with lattice data, is taken from Ref. \cite{Z}, and $\alpha$=0.9 is
adopted from Ref. \cite{TongZhao}.

\begin{figure}
\includegraphics[width=3.5in]{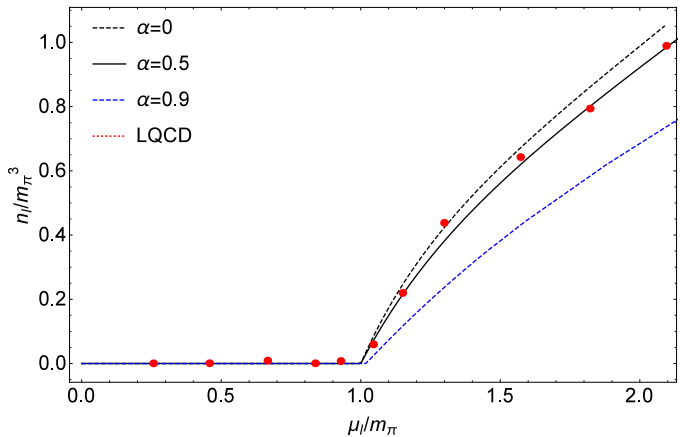}
\caption{The normalized isospin density $n_I/m_{\pi}^3$ as a function of the normalized isospin chemical potential $\mu_I/m_{\pi}$ at $T =$ $\mu_B$ $= 0$.}
\label{s3 1}
\end{figure}

\begin{figure}
\includegraphics[width=3.5in]{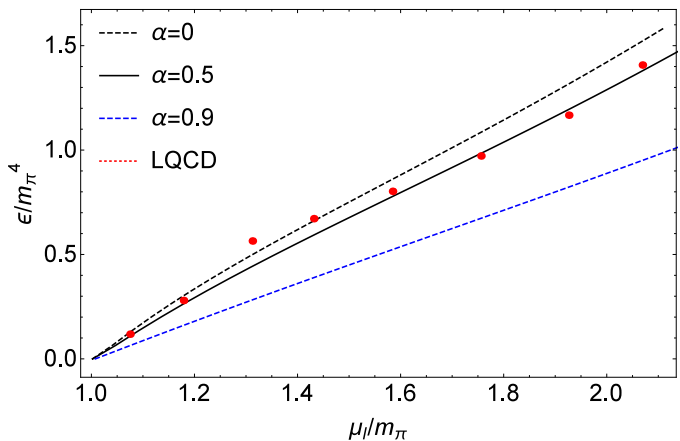}
\caption{The normalized energy density $\epsilon/m_{\pi}^4$ as a function of the normalized isospin chemical potential $\mu_I/m_{\pi}$ at $T =$ $\mu_B$ $= 0$.}
\label{s3 2}
\end{figure}

\begin{figure}
\includegraphics[width=3.5in]{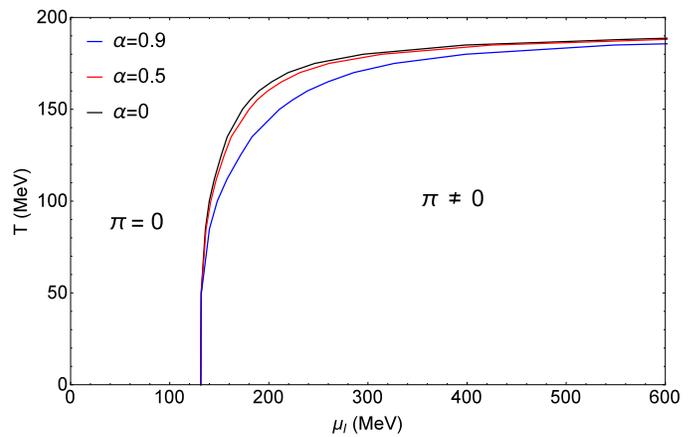}
\caption{The phase diagram of pion superfluidity in $T-\mu_I$ plane $(\mu_B=0)$.}
\label{s3 3}
\end{figure}

Figs. \ref{s3 1} and \ref{s3 2} show the variation of the normalized isospin density and energy density with $\alpha=0,0.5,0.9$ versus $\mu_I$
scaled by $m_{\pi}$ respectively \cite{Z}. We can realize that the lattice QCD data can be well described by the calculation with $\alpha=0.5$,
only some data are situated on the $\alpha=0$ curve (i.e. the standard NJL model results) around $\mu_I \sim 1.5m_{\pi}$. The solutions of
Eq. (\ref{s2 5}) for $\pi$ condensate with different $\alpha$'s delimitating the regions of the pion superfluidity phase ($\pi\ne 0$) and
the normal phase ($\pi=0$) are given in Fig. \ref{s3 3} in the $T-\mu_I$ plane for $\mu_B=0$. It can be seen that the phase transition line
of these two regions will become lower as $\alpha$ increases (the $\alpha=0$ corresponding to the result of Ref. \cite{Lian}) except
in the beginning they always keep zero, and the largest differences among them appear at $\mu_I \sim 1.5m_{\pi}$. Especially at zero temperature
the critical isospin chemical potential $\mu_I^C$ of the phase transition all have $\mu_I^C=m_{\pi}$.

\begin{figure}
\includegraphics[width=3.5in]{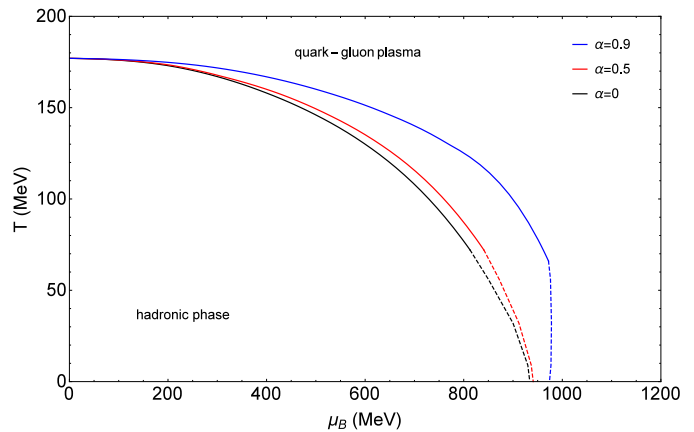}
\caption{The phase diagram of chiral condensate in $T-\mu_B$ plane $(\mu_I=0)$.}
\label{s3 4}
\end{figure}

In the following, we will discuss the phase diagram at finite baryon chemical potential for different fixed isospin chemical potentials.
By solving the Eqs. (\ref{s2 9})-(\ref{s2 11}) numerically we can obtain the phase diagram (Fig. \ref{s3 4}) in $T-\mu_B$ plane at $\mu_I=0$.
The chiral phase transition lines with different $\alpha$'s in the diagram represent the phase transition from hadronic phase to QGP phase,
in which the solid lines denote the crossover and the dashed lines denote the first-order phase transition. We can see that for the same
temperature the critical baryon chemical potential of the phase transition will get large with $\alpha$ increasing, which means that at a
fixed temperature for the larger $\alpha$ the occurrence of the phase transition will be postponed as baryon chemical potential increases.

\begin{figure}
\includegraphics[width=3.5in]{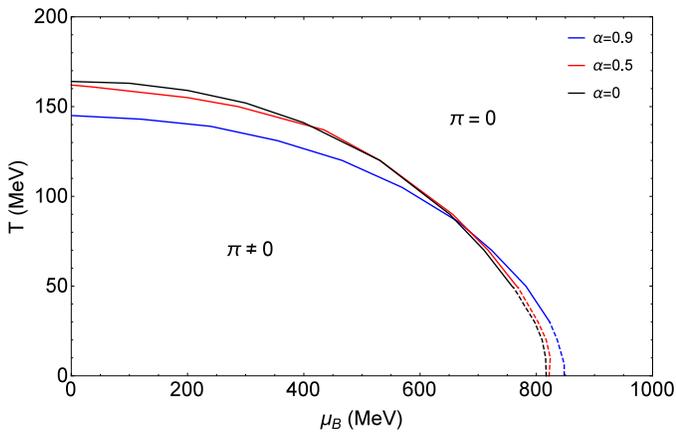}
\caption{The phase diagram of pion superfluidity in $T-\mu_B$ plane ($\mu_I=200$ $\text{MeV}$).}
\label{s3 5}
\end{figure}

\begin{figure}
\includegraphics[width=3.5in]{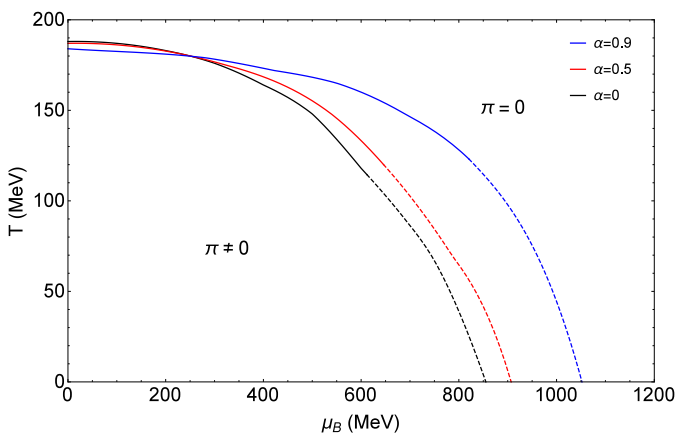}
\caption{The phase diagram of pion superfluidity in $T-\mu_B$ plane ($\mu_I=500$ $\text{MeV}$).}
\label{s3 6}
\end{figure}

We solve the Eqs. (\ref{s2 4}) and (\ref{s2 5}) simultaneously and obtain the phase diagrams (Figs. \ref{s3 5} and \ref{s3 6}) in $T-\mu_B$ plane
at $\mu_I=200$ $\text{MeV}$ and $\mu_I=500$ $\text{MeV}$. There is an onset of pion condensation (i.e. $\pi\neq 0$ corresponding to the system
in the pion superfluidity phase) at $\mu_I^C=m_{\pi}$ and $T=\mu_B=0$  \cite{D.T.Son}. From the figures, we can see that at low temperatures
the system transiting from pion superfluidity phase to normal phase ($\pi=0$) needs much large $\mu_B$. As shown in Figs. \ref{s3 5} and \ref{s3 6},
the solid lines indicate the second-order phase transition, the dashed lines indicate the first-order phase transition, and for the lower $\mu_I=200$ MeV the phase transition lines with different $\alpha$'s (the $\alpha=0$ corresponding to the result of Ref. \cite{Lian}) intersect at low $T$
and high $\mu_B$ while for the higher $\mu_I=500$ MeV they intersect at high $T$ and low $\mu_B$.

\begin{table*}
\centering
\caption{The critical endpoint (CEP) with different $\alpha$'s at $\mu_I=0$ and the tricritical point (TCP) with different $\alpha$'s at $\mu_I=200,500$ $\text{MeV}$.}
\label{s4_10}
\begin{tabular}{cccc}
\toprule
            ~~& $\alpha=0$ & $\alpha=0.5$ & $\alpha=0.9$ \\
\colrule
$\mu_I= 0$   & $(T=72 $\text{MeV}$, \mu_B=813 $\text{MeV}$)$ & $(T=72 $\text{MeV}$, \mu_B=841 $\text{MeV}$)$ & $(T=69 $\text{MeV}$, \mu_B=971 $\text{MeV}$)$ \\
$\mu_I= 200$ $\text{MeV}$ & $(T=49 $\text{MeV}$, \mu_B=761 $\text{MeV}$)$ & $(T=49 $\text{MeV}$, \mu_B=769 $\text{MeV}$)$ & $(T=30 $\text{MeV}$, \mu_B=823 $\text{MeV}$)$ \\
$\mu_I= 500$ $\text{MeV}$ & $(T=111 $\text{MeV}$, \mu_B=612 $\text{MeV}$)$ & $(T=120 $\text{MeV}$, \mu_B=646 $\text{MeV}$)$ & $(T=123 $\text{MeV}$, \mu_B=821 $\text{MeV}$)$ \\
\botrule
\end{tabular}
\label{s3 10}
\end{table*}

At $\mu_I=0$, the phase diagram Fig. \ref{s3 4} of chiral condensate gives the critical endpoint (CEP) with different $\alpha$'s, which is
terminal point of the first-order phase transition curve. In the cases of $\mu_I=200,500$ $\text{MeV}$, the phase diagrams
Fig. \ref{s3 5} and Fig. \ref{s3 6} of pion superfluidity give the tricritical point (TCP) with different $\alpha$'s separately, which is
the intersection of the first-order phase transition curve and the second-order phase transition curve. In Table \ref{s3 10},
the critical points of phase transition are presented. From table we can see that for CEPs the larger the value of $\alpha$ the larger
the critical baryon chemical potential but the smaller the critical temperature; for TCPs at the lower fixed isospin chemical potential
$\mu_I=200$ $\text{MeV}$ the variation of the critical baryon chemical potential and the critical temperature versus $\alpha$ is similar
to the case of CEPs, but at the higher fixed isospin chemical potential $\mu_I=500$ $\text{MeV}$ the larger the value of $\alpha$
the larger the critical temperature.

\begin{figure}
\includegraphics[width=3.5in]{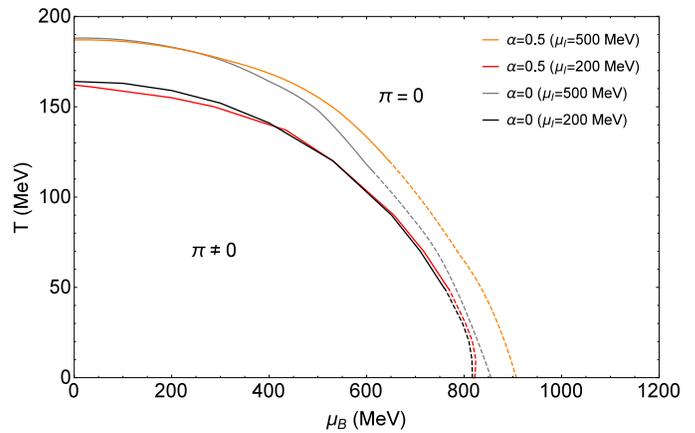}
\caption{The phase diagram of pion superfluidity in $T-\mu_B$ plane ($\mu_I=200,500$ $\text{MeV}$).}
\label{s3 7}
\end{figure}

\begin{figure}
\includegraphics[width=3.5in]{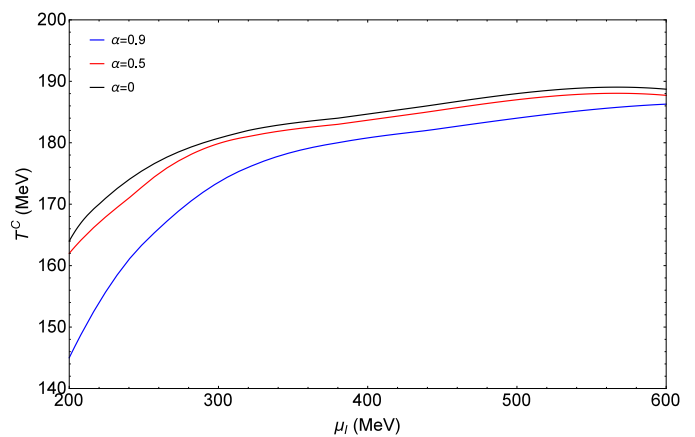}
\caption{The critical temperature $T^C$ of phase transition changes with
$\mu_I$ at $\mu_B=0$.}
\label{s3 8}
\end{figure}

\begin{figure}
\includegraphics[width=3.5in]{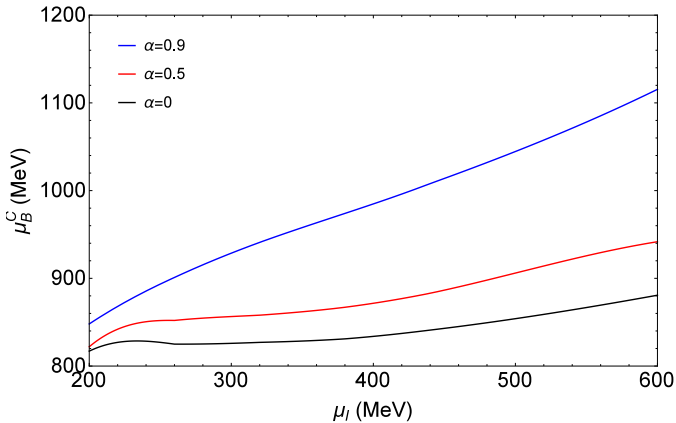}
\caption{The critical baryon chemical potential $\mu_B^C$ of phase transition changes with
$\mu_I$ at $T=0$.}
\label{s3 9}
\end{figure}

As plotted in Fig. \ref{s3 7}, we compare the results with $\alpha=0.5$ (fits well with Lattice data) and the results with $\alpha=0$
(the standard NJL model results) for different fixed nonzero isospin chemical potentials. The figure indicates that for
the same $\alpha$ the increasing $\mu_I$ moves the phase transition line to its upper right in the $T-\mu_B$ plane, and for
the $\alpha=0$ case the part of the first-order phase transition line move less.

We display the critical temperature $T^C$ of phase transition with different $\alpha$'s at $\mu_B=0$ as a function of $\mu_I$ in Fig. \ref{s3 8}.
It shows that the critical temperature $T^C$ of phase transition with different $\alpha$ increases quickly at the beginning as the isospin
chemical potential increases, and when $\mu_I > 300$ MeV the critical temperature $T^C$ increases slowly to a constant value.
And on the whole the $T^C$ values are closer between $\alpha=0$ and $\alpha=0.5$ compared with $\alpha=0.9$.
The critical baryon chemical potential $\mu_B^C$ of phase transition with different $\alpha$'s at $T=0$ as a function of $\mu_I$ is plotted
in Fig. \ref{s3 9}. For small $\mu_I$, the values of $\mu_B^C$ with different $\alpha$ are closer, and as the isospin chemical potential increases
the $\mu_B^C$ with $\alpha=0, 0.5$ increase slowly and synchronously, while the $\mu_B^C$ with $\alpha=0.9$ almost increases linearly.

\section {Summary}
\label{s4}
In this paper, the self-consistent mean field approximation of the NJL model is employed to study QCD phase structure
at finite densities and finite temperature. In our calculation, we consider the contributions of different interaction channels (the introduced parameter $\alpha$ reflects the weight) with different cases of $\alpha=0$, $\alpha=0.5$, and $\alpha=0.9$, where the results with $\alpha=0.5$ are in good agreement with lattice data \cite{Z}. We found that the phase transition line of the $T-\mu_I$ plane with $\alpha=0.5$ is located in the middle in the three cases of $\alpha=0,0.5,0.9$, and the difference of these three lines is largest at $\mu_I \sim 1.5m_{\pi}$.

We plot the phase transition lines in the $T-\mu_B$ plane for different fixed isospin chemical potentials. At $\mu_I=0$, the chiral phase diagram Fig. \ref{s3 4} gives the CEPs with different $\alpha$'s and it can be seen that the critical baryon chemical potential of the CEP will increase with $\alpha$ increasing. At $\mu_I=200$ $\text{MeV}$ and $500$ $\text{MeV}$ ($>$ $m_\pi$), the corresponding phase diagrams Figs. \ref{s3 5} and \ref{s3 6} of pion condensate give the TCPs with different $\alpha$'s, and we can see that at low temperatures the system is in the pion superfluidity phase at first and then it transits to the normal phase with $\mu_B$ increasing, and the smaller critical baryon chemical potential of the TCP is obtained for higher fixed $\mu_I$.

Finally, the critical temperature $T^C$ and the critical baryon chemical potential $\mu_B^C$ of phase transition change with
$\mu_I$ are respectively discussed. It finds that for $\mu_I > 300$ MeV the variation rate of the critical temperature $T^C$ of phase transition with different $\alpha$'s will decease as $\mu_I$ increases and the critical baryon chemical potential $\mu_B^C$ of phase transition with $\alpha=0, 0.5$ will rise slowly as $\mu_I$ increases compared with $\alpha=0.9$. In general, through investigating QCD phase structure and the location of critical point we can further understand the properties of phase transition at finite densities and finite temperature, and this also provides some information of the phase transition for the measurement of the experiment.

\acknowledgments
This work are supported by National Natural Science Foundation of China (under Grants No. 12075117, No. 11535005, 11775118, 11690030 and No. 11905104) and National Major state Basic Research and Development of China (2016YEF0129300).

\end{document}